\documentclass[aps,prb,superscriptaddress,twocolumn,showpacs,amsmath,amssymb]{revtex4}
\usepackage{graphicx}
\usepackage{bm}

\begin{document}


\title{Ni-impurity effects on the superconducting gap of La$\rm _{2-\it x}$Sr$_{x}$CuO$_4$ studied from the magnetic field and temperature dependence of the electronic specific heat}


\author{T.~Kurosawa}
\affiliation{Department of Physics, Hokkaido University, Sapporo
060-0810, Japan}

\author{N.~Momono}%
\affiliation{Department of Applied Sciences,
Muroran Institute of Technology, Muroran 050-8585, Japan}

\author{M.~Oda}
\affiliation{Department of Physics, Hokkaido University, Sapporo
060-0810, Japan}

\author{M.~Ido}
\affiliation{Department of Physics, Hokkaido University, Sapporo
060-0810, Japan}


\date{\today}

\begin{abstract}
The magnetic field and temperature dependence of the electronic specific heat $C_{\rm el}$ have been systematically investigated in $\rm La_{2-{\it x}}Sr_{\it x}Cu_{1-{\it y}}Ni_{\it y}O_4$ (LSCNO) in order to study Ni-impurity effects on the superconducting (SC) gap. In LSCNO with $x$=0.15 and $y$=0.015, the value of $\gamma$ ($\equiv C_{\rm el}/T$) at $T$=0 K, $\gamma_0$, is enhanced under the magnetic field $H$ applied along the $\bm c$-axis. The increment of $\gamma_0$, $\Delta \gamma_0$, follows the Volovik relation $\Delta \gamma_0$=$A\sqrt{H}$, characteristic of the SC gap with line nodes, with prefactor $A$ similar to that of a pure sample. The $C_{\rm el}/T$ vs. $T$ curve under $H$=0 shows a $d$-wave-like SC anomaly with an abrupt increase at $T_{\rm c}$ and $T$-linear dependence at $T$$\ll$$T_{\rm c}$, although the $\gamma_0$-value in the $C_{\rm el}/T$ vs. $T$ curve increases with increasing Ni concentrations. Interestingly, as the SC part of $C_{\rm el}/T$, $C_{\rm el}/T$$-$$\gamma_0$$\equiv$$\gamma_{\rm s}$, decreases in LSCNO, $T_{\rm c}$ is reduced in proportion to the decrease of $\gamma_{\rm s}$. These findings can be explained phenomenologically by a simple model in which Ni impurities bring about strong pair breaking at the edges of the coherent nodal part of the Fermi surface but in the vicinity of the nodes of the SC gap. The reduction of the SC condensation energy $U_0$ in LSCNO, evaluated from $C_{\rm el}$ at $T$\hspace{0.3em}\raisebox{0.4ex}{$<$}\hspace{-0.75em}\raisebox{-.7ex}{$\sim$}\hspace{0.3em}$T_{\rm c}$, is also understood by the same model.
\end{abstract}

\pacs{74.62.En, 74.72.-h,}

\maketitle

\section{Introduction}
One of the striking features in high-$T_{\rm c}$ cuprates is an unusual impurity effect on the superconductivity, which has been studied extensively from experimental and theoretical points of view to elucidate the mechanism for high-$T_{\rm c}$ superconductivity. As is well known, in conventional superconductors, whose superconducting (SC) order parameter is of the $s$-wave type, magnetic impurities strongly suppress $T_{\rm c}$\cite{1Abrikosov}, whereas nonmagnetic ones do little\cite{2Anderson}. However, this is not the case for the impurity effect on high-$T_{\rm c}$ superconductivity, whose order parameter is of the $d$-wave type. In high-$T_{\rm c}$ cuprates, a small number of in-plane impurities, such as nonmagnetic Zn or magnetic Ni impurities, which are partially substituted for Cu sites within CuO$_2$ planes, cause a large reduction of $T_{\rm c}$ and a large increase of the residual density of states (DOS) at the Fermi level ($E_{\rm F}$), whether they are magnetic or not.\cite{3Tarascon, 4Xiao, 5Koike, 6Ishida, 7Nakano, 8Adachi} More interestingly, it has been demonstrated in $\rm La_{2-{\it x}}Sr_{\it x}CuO_4$ (LSCO) that a nonmagnetic Zn impurity exhibits a stronger effect on the high-$T_{\rm c}$ superconductivity than a magnetic Ni impurity; in particular, the recovery of the residual DOS($E_{\rm F}$) is much more evident for the former impurity.\cite{6Ishida, 7Nakano, 8Adachi} The large recovery of the residual DOS($E_{\rm F}$) due to a small amount of Zn impurity has been explained well in terms of a model in which the SC gap is completely suppressed locally in areas with a diameter on the order of the antiferromagnetic (AF) correlation length around the Zn impurity, where the AF correlation among Cu spins is disturbed because there are no localized spins on the impurity sites.\cite{7Nakano, 9Nachumi} However, such a model gives no explanation for the large suppression of $T_{\rm c}$ in the SC region outside the magnetically disturbed areas around Zn impurities. Furthermore, the mechanism for the suppression of superconductivity in Ni-doped LSCO has not been clarified yet. It has been demonstrated in measurements of the $T$ dependence of magnetic susceptibility in pure and Ni-doped LSCO that a small amount of Ni impurity, which has a localized spin, has little effect on the AF correlation among Cu-spins, as in the SC region outside the non-SC areas around Zn impurities.\cite{7Nakano, 10Hiraka} Therefore, the suppression of superconductivity in Ni-doped LSCO is expected to occur rather uniformly over the whole crystal, which is much simpler than that in Zn-doped LSCO. 

In $d$-wave superconductors, it has been broadly thought that the presence of impurities gives rise to pair breaking in the vicinity of the nodes and leads to collapse of the SC gap there.\cite{11Haas} However, in our recent study on the low-temperature ($T$$\ll$$T_{\rm c}$) electronic specific heat of Ni-doped LSCO, we pointed out that the SC gap around the nodes will be affected only slightly by a small amount of Ni impurity.\cite{12Kurosawa} Furthermore, recent angle-resolved photoemission spectroscopy (ARPES) experiments on $\rm Bi_2Sr_{2-{\it x}}{\it R}_{\it x}CuO_{6+\delta }$ ($R$-Bi2201, $R$: rare earth element) reported that as interplane disorders, which are caused by the substitution of $R$ atoms for Sr sites, are strengthened by replacing the $R$ atom La with Eu, the $T_{\rm c}$ is reduced from 33 K to 18 K at the optimal doping level, but the SC gap dispersion around the nodes remains almost unchanged. The suppression of superconductivity in Eu-Bi2201 has been discussed in terms of the shrinkage of a nodal part of the Fermi surface (FS), the so-called ``Fermi arc'', which consists of coherent electronic states and is responsible for an effective SC gap in determining $T_{\rm c}$.\cite{13Okada} On the other hand, it has been claimed in ARPES experiments on $\rm Bi_2Sr_2CaCu_{2-\it x}{\it M}_{\it x}O_{8+\delta }$ ($M$-Bi2212, $M$=Ni or Zn) that the SC gap is collapsed in the vicinity of the nodes. However, its region around the nodes is only $\sim$2\% of the whole FS, corresponding to the region whose original gap size is smaller than $\sim$1 meV, which would be too small to explain the observed reduction of $T_{\rm c}$ ($\Delta T_{\rm c}$=5$-$10 K).\cite{14Sato} Thus, in high-$T_{\rm c}$ cuprates, the impurity or disorder effect on the SC gap is still under debate even for the nodal region.

One of the good ways to investigate the SC gap around the nodes in $d$-wave superconductors is to measure the electronic specific heat $C_{\rm el}$ in the SC mixed state induced by the application of magnetic field $H$ along the lines of nodes. In the mixed state, the electronic specific heat coefficient $\gamma$ ($\equiv$$C_{\rm el}/T$) at $T$=0 K, the residual $\gamma$ value, $\gamma_0$, reflecting the residual DOS($E_{\rm F}$), increases with the increase of $H$. This is because the residual DOS($E_{\rm F}$) recovers in areas extending along the node directions from vortex cores, which is due to the Doppler shift of the quasiparticle dispersion curve. The increment of $\gamma_0$, $\Delta \gamma_0$, follows the Volovik relation, $\Delta \gamma_0$=$A\sqrt{H}$, where prefactor $A$ is inversely proportional to the slope $v_{\Delta}$ of the gap dispersion around the nodes.\cite{15Volovik} The Volovik relation has been confirmed in pure samples of $\rm YBa_2Cu_3O_7$\cite{16Moler} and LSCO\cite{17Nohara, 18Wen}.

In the present study, the Ni-impurity effect on the SC gap was examined from measurements of the electronic specific heat $C_{\rm el}$ at $T$$\ll$$T_{\rm c}$ under magnetic fields for $\rm La_{2-{\it x}}Sr_{\it x}Cu_{1-{\it y}}Ni_{\it y}O_4$ (LSCNO) samples. First, we report that the residual $\gamma$ value ($\equiv$$C_{\rm el}/T$) at $T$=0 K, $\gamma_0$, increases under magnetic fields, following the Volovik relation, in LSCNO as well as in pure samples. Next we report that the SC part of $C_{\rm el}/T$, $C_{\rm el}(T, y)/T$$-$$\gamma_0(y)$$\equiv$$\gamma_{\rm s}(T, y)$, under $H$=0 shows a SC anomaly exhibiting an abrupt increase at $T_{\rm c}$ and $T$-linear dependence at $T$$\ll$$T_{\rm c}$, although the SC anomaly is largely suppressed as a whole with increases in Ni-concentration $y$. Furthermore, $T_{\rm c}$ is reduced in proporion to the suppression of $\gamma_{\rm s}(T, y)$ in LSCNO. On the basis of these observations, we propose a scenario in which the SC gap of LSCNO will be predominantly suppressed not in the vicinity of the nodes but at the edges of the coherent nodal FS. Such Ni-impurity pair breaking leads to natural explanations for the marked suppression of $T_{\rm c}$ and the SC condensation energy $U_0$, evaluated from the data of $\gamma_{\rm s}(T, y)$, in LSCNO.

\section{Experimental Procedures}
Specific heat measurements at low temperatures ($T$$<$10 K) under magnetic fields, whose directions were perpendicular to CuO$_2$ planes, that is, along the $\bm c$-axis, were carried out by a thermal relaxation method in single crystal samples of LSCNO. On the other hand, specific heat measurements in a wide $T$ range from 4 K to 60 K, sufficiently higher than $T_{\rm c}$, under a zero magnetic field were carried out by a conventional heat-pulse technique on ceramic samples of LSCNO. The way to determine the $T$ dependence of the phonon term $C_{\rm ph}(T)$ in a wide $T$ range is one of the important keys to extracting the change of $C_{\rm el}(T)$ in accordance with the SC transition from the total specific heat $C(T)$, including both the $C_{\rm el}(T)$ and $C_{\rm ph}(T)$ terms. The partial Ni substitution for Cu sites makes it possible to determine the phonon term $C_{\rm ph}(T)$ of the SC samples. This is because a small amount of Ni substitution can destroy the superconductivity completely but has little effect on the phonon term, at least, in the optimal samples.\cite{7Nakano} This Ni substitution effect is in sharp contrast with Zn substitution, which modifies the phonon term significantly. In the present study, we obtained the electronic term $C_{\rm el}(T)$ by subtracting the phonon term $C_{\rm ph}^{\rm Ni}(T)$, which was estimated in a Ni-doped non-SC sample, from the observed, total specific heat $C(T)$; $C_{\rm el}(T)$=$C(T)$$-$$C_{\rm ph}^{\rm Ni}(T)$. Details of the estimations of $C_{\rm ph}^{\rm Ni}(T)$ and/or $C_{\rm el}(T)$ have been published elsewhere.\cite{19Nagata, 20Momono, 21Matsuzaki}

Ceramic and single-crystal samples of LSCNO were prepared by using La$_2$O$_3$, SrCO$_3$, CuO, and NiO powders of 99.99-99.999\% purity. For the ceramic samples, these powders were mixed well and then heat-treated in a furnace at a certain temperature under flowing oxygen. The heat-treated materials were reground well and pressed into pellets, and were sintered in flowing oxygen. The single crystal samples were grown by the traveling solvent floating zone method. For these ceramic and single crystal samples, the SC critical temperature $T_{\rm c}$ was determined from the SC diamagnetism measured with a SQUID magnetometer. 

\section{Results and Discussion}

\subsection{Electronic specific Heat in the Mixed State of La$_{2-{\it x}}$Sr$_{\it x}$Cu$_{1-{\it y}}$Ni$_{\it y}$O$_4$}
\begin{figure}
\begin{center}
\includegraphics[scale=.50]{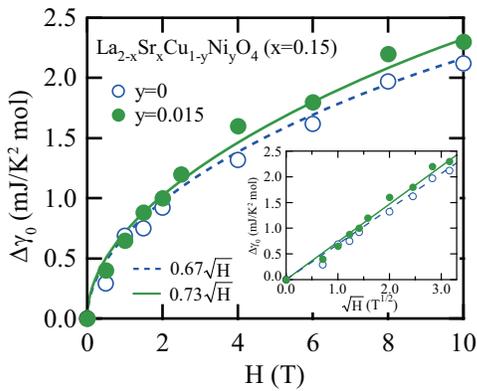}
\end{center}
\caption{(Color online) Magnetic field dependence of $\Delta \gamma_0$=$\gamma_0(H)$$-$$\gamma_0(0)$ in pure $\rm La_{2-{\it x}}Sr_{\it x}CuO_4$ ($x$=0.15, $T_{\rm c}$=37.0 K) (open circles) and that of $\rm La_{2-{\it x}}Sr_{\it x}Cu_{1-{\it y}}Ni_{\it y}O_4$ ($x$=0.15, $y$=0.015, $T_{\rm c}$=26.5 K) (closed circles) samples. The solid and dashed lines represent functions of $\sqrt{H}$. The inset shows $\Delta \gamma_0$ vs. $\sqrt{H}$ plots.}
\label{fig2}
\end{figure}

In $\rm La_{2-{\it x}}Sr_{\it x}Cu_{1-{\it y}}Ni_{\it y}O_4$ (LSCNO) with $x$=0.15, the $C/T$ vs. $T^2$ curve at $H$=0 follows a straight line at $T$$\ll$$T_{\rm c}$, because the electronic $T^2$ term arising from low-energy quasiparticle excitations around the line nodes at $T$$\ll$$T_{\rm c}$ is much smaller than the phonon $T^3$ term except the nearly over-doped region.\cite{22Momono, 23Momono} Applying a magnetic field to the pure crystal with $y$=0 along the direction perpendicular to the CuO$_2$ plane (the $\bm c$-axis), the $C/T$ vs. $T^2$ curve shifts upward as a whole, indicating that the value of $\gamma_0$ is enhanced under magnetic fields through the Doppler shift of the quasiparticle dispersion curve, as shown in Fig. 1(a) in ref. (12). In the present study, we obtained the $H$ dependence of $\gamma_0$, $\gamma_0(H)$, up to 10 T more precisely than in our previous work (Fig.\ \ref{fig2}). One can see that the increment of $\gamma_0$, $\Delta \gamma_0$ [$\equiv$$\gamma_0(H)$$-$$\gamma_0(0)$], follows the Volovik relation $\Delta \gamma_0$=$A\sqrt{H}$, as has been reported for pure high-$T_{\rm c}$ cuprates\cite{16Moler, 17Nohara, 18Wen}.

Prefactor $A$ in the Volovik relation is written as\cite{24Hussey}
\begin{equation}
A = \frac{4k_{\rm B}^2}{3\hbar}\sqrt{\frac{\pi}{\Phi_0}}\frac{nV_{\rm mol}}{l_c}\frac{\eta}{v_{\rm \Delta}}. \label{eq:volovik}
\end{equation}
In this expression, $v_{\Delta}$ is the slope of the $d$-wave gap dispersion around the nodes, $n$ the number of CuO$_2$ planes within the unit cell, $l_c$ the lattice constant along the $\bm c$-axis, and $V_{\rm mol}$ the volume of a mole. Prefactor $A$ also depends on the parameter $\eta$; $\eta$ is $\eta$=0.5 for a square flux-line lattice and $\eta$=0.47 for a triangular one. Small angle neutron scattering experiments on nearly optimally doped and overdoped LSCO crystals under magnetic fields parallel to the $\bm c$-axis have demonstrated that the vortex lattice is of a triangular type for low fields and gradually changes into a square type around $H$$\sim$0.5 T.\cite{25Gilardi} The magnetic fields (0.5 T$\le$$H$$\le$10 T) used in the present study on LSCNO with $x$=0.15 were in the $H$ range of the square-type lattice. Thus, using the experimental value of $A$ and $\eta$=0.5, $v_{\Delta}$ is estimated to be 6.1$\times$10$^5$ cm/s from Eq.\ \ref{eq:volovik}. This value for $v_{\Delta}$ is in good agreement with that obtained in ARPES experiments on LSCO crystals with $x$$\sim$0.15.\cite{26Terashima, 27Shi, 28Yoshida}

The $H$ dependence of $C/T$ vs. $T^2$ curves for LSCNO with $x$=0.15 and $y$=0.015, whose $T_{\rm c}$ (=26.5 K) is $\sim$30$\%$ lower than that of the pure crystal is shown in Fig. 1(b) in ref. (12). One can see in this figure that the $C/T$ vs. $T^2$ curve also shifts upward as a whole under magnetic fields; that is, the value of $\gamma_0$ is enhanced in the LSCNO sample by the Doppler shift of the quasiparticle dispersion curve. We estimate the $\gamma_0$ value under magnetic fields from $C/T$ vs. $T^2$ curves of LSCNO in the same way as in the pure crystal, and plot the increment of $\gamma_0$, $\Delta \gamma_0$ [=$\gamma_0(H)$$-$$\gamma_0(0)$], as a function of $H$ in Fig.\ \ref{fig2}. The $\Delta \gamma_0$ thus estimated also satisfies the Volovik relation $\Delta \gamma_0$=$A\sqrt{H}$, although prefactor $A$ in the present LSCNO is slightly larger than that in the pure crystal. The maintenance of the Volovik relation means that nodes remain in the SC gap in LSCNO as if they are almost free from the pair breaking with Ni impurity. According to Eq.\ \ref{eq:volovik}, the present increment of $A$ leads to a 9\% reduction of $v_\Delta$. Since $v_\Delta$ is proportional to the gap maximum $\Delta_0$ (see Eq.\ \ref{eq:volovik}), $T_{\rm c}$ is expected to decrease in accordance with the reduction of $v_{\Delta}$. However, in the present experimental result, the 9\% reduction of $v_\Delta$ is too small to explain the $\sim$30\% reduction of $T_{\rm c}$.
\label{sec:sec31}

\subsection{Temperature dependence of the electronic specific heat {\it C}$_{\rm el}$ in La$_{2-{\it x}}$Sr$_{\it x}$Cu$_{1-{\it y}}$Ni$_{\it y}$O$_4$}
\begin{figure}
\begin{center}
\includegraphics[scale=.50]{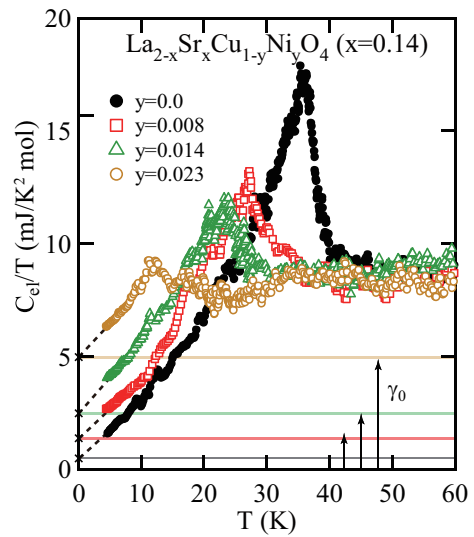}
\end{center}
\caption{(Color online) $C_{\rm el}(T)/T$ vs. $T$ plots for $\rm La_{2-{\it x}}Sr_{\it x}Cu_{1-{\it y}}Ni_{\it y}O_4$ with $x$=0.14. The Ni concentration $y$ and the corresponding $T_{\rm c}$ are as follows: $y$=0 ($T_{\rm c}$=36.8 K), $y$=0.008 ($T_{\rm c}$=28.8 K), $y$=0.014 ($T_{\rm c}$=24.3 K), and $y$=0.023 ($T_{\rm c}$=13.0 K).}
\label{fig3}
\end{figure}

Figure\ \ref{fig3} shows the electronic specific heat $C_{\rm el}(T)/T$ over a wide $T$ range from 4 K to 60 K for slightly underdoped LSCNO ceramics ($x$=0.14, $y$=0, 0.008, 0.014, and 0.023). In the pure sample ($y$=0), the $C_{\rm el}(T)/T$ vs. $T$ curve with a very small $\gamma_0$ exhibits a SC anomaly with an abrupt increase around $T_{\rm c}$ and a $T$-linear dependence at $T$$\ll$$T_{\rm c}$. The $T$-linear dependence at $T$$\ll$$T_{\rm c}$ originates in the existence of the line nodes in the SC gap. In LSCNO samples, the $\gamma_0$ value in the $C_{\rm el}(T)/T$ vs. $T$ curve is largely enhanced, meaning that parts of the FS revive on account of pair breaking caused by Ni impurity (Fig.\ \ref{fig3}). As $\gamma_0$ is enhanced with the increase of the Ni-impurity concentration $y$, the SC anomaly shifts toward low temperatures as a whole, and is suppressed with the increase of $y$, as seen in Fig.\ \ref{fig3}. 

\begin{figure}
\begin{center}
\includegraphics[scale=.50]{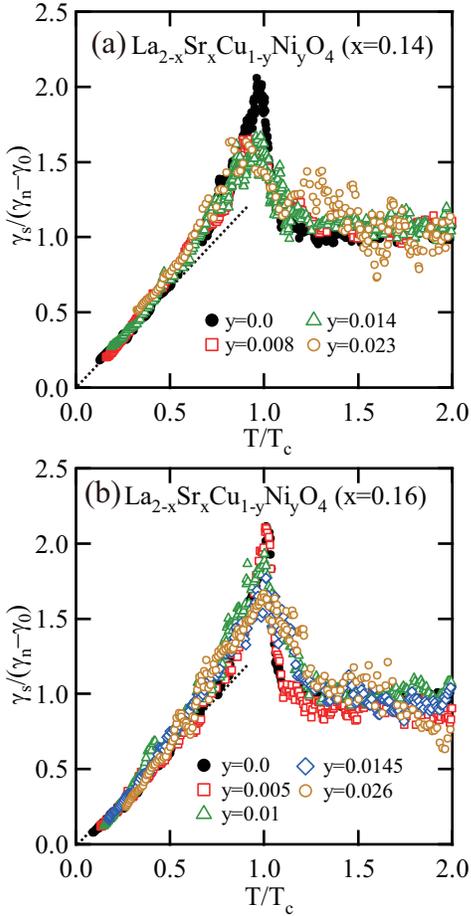}
\end{center}
\caption{(Color online) $\gamma_{\rm s}$/($\gamma_{\rm n}$$-$$\gamma_0$) vs. $T/T_{\rm c}$ plots for $\rm La_{2-{\it x}}Sr_{\it x}Cu_{1-{\it y}}Ni_{\it y}O_4$ samples with (a) $x$=0.14 and (b) $x$=0.16.}
\label{fig5}
\end{figure}

\begin{figure}
\begin{center}
\includegraphics[scale=.50]{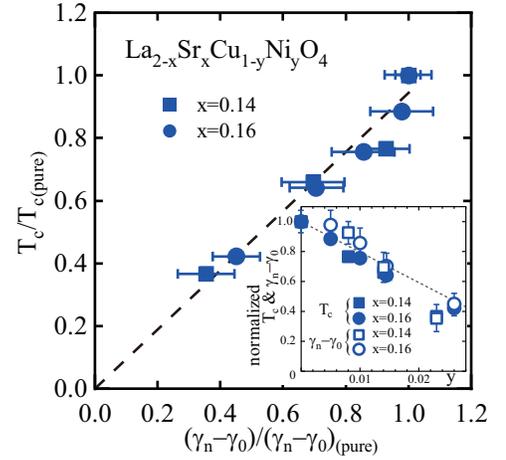}
\end{center}
\caption{(Color online) $T_{\rm c}/T_{\rm c(pure)}$ vs. ($\gamma_{\rm n}$$-$$\gamma_0$)/($\gamma_{\rm n}$$-$$\gamma_0$)$_{\rm (pure)}$ plots for $\rm La_{2-{\it x}}Sr_{\it x}Cu_{1-{\it y}}Ni_{\it y}O_4$ with $x$=0.14 and $x$=0.16. The inset shows the $y$ dependence of $T_{\rm c}$ and $\gamma_{\rm n}$$-$$\gamma_0$ value normalized with those of the pure sample, $T_{\rm c(pure)}$ and ($\gamma_{\rm n}$$-$$\gamma_0$)$_{\rm (pure)}$, respectively.}
\label{fig6}
\end{figure}

Here we pay attention to the SC part of $C_{\rm el}(T, y)/T$, namely, $C_{\rm el}(T, y)/T$$-$$\gamma_0(y)$. We normalize the SC part $C_{\rm el}(T, y)/T$$-$$\gamma_0(y)$$\equiv$$\gamma_{\rm s}(T, y)$ with its normal state value $\gamma_{\rm n}(y)$$-$$\gamma_0(y)$, and plot it as a function of $T/T_{\rm c}(y)$ in Fig.\ \ref{fig5}(a) for $x$=0.14 and \ref{fig5}(b) for $x$=0.16. The normalized SC parts are in agreement with each other, including that of the pure sample, though the jump at $T_{\rm c}$ is smeared in some degree in LSCNO, as seen in Fig.\ \ref{fig5}. Such agreement indicates that $\gamma_{\rm n}(y)$$-$$\gamma_0(y)$ reflects the fraction of the quasiparticles which become SC condenstates at $T$$\ll$$T_{\rm c}$. Actually, $\gamma_{\rm n}(y)$$-$$\gamma_0(y)$, which largely decreases with increasing $y$, closely relates to the rapid suppression of $T_{\rm c}(y)$ in LSCNO, as shown in Fig.\ \ref{fig6}. The close relation between $\gamma_{\rm n}(y)$$-$$\gamma_0(y)$ and $T_{\rm c}(y)$ is expected to provide a key to understanding the Ni-impurity effect. It is also worth noting that the normalized $\gamma_{\rm s}(T, y)$ vs. $T$ plot shows the $T$-linear dependence at $T$$\ll$$T_{\rm c}(y)$ in all samples, including the pure sample, as seen in Fig.\ \ref{fig5}. This result implies that the line nodes remain in the SC gap of LSCNO, which is consistent with the present findings in specific heat measurements under magnetic fields. 

On the basis of the above results and discussion, we propose the following simple scenario for the Ni-impurity effect on the SC gap. The SC gap is suppressed predominantly on segments of the FS away from the nodes, but remains virtually unchanged in the vicinity of the nodes. The former segments of the FS, where the SC gap is predominantly suppressed, lead to the $T$-independent part $\gamma_0(y)$ in the $C_{\rm el}(T, y)/T$ vs. $T$ curve through the revival of the DOS($E_{\rm F}$) resulting from the local suppression of the SC gap. On the other hand, the latter nodal segments are SC segments and responsible for the SC part $\gamma_{\rm s}(T, y)$. In such a case, the length $L(y)$ of the SC nodal segment is proportional to $\gamma_{\rm n}(y)$$-$$\gamma_0(y)$; $L(y)$$\propto$$\gamma_{\rm n}(y)$$-$$\gamma_0(y)$. On the basis of the relation $L(y)$$\propto$$\gamma_{\rm n}(y)$$-$$\gamma_0(y)$ and the coherent SC nodal segment (the Fermi arc), we will discuss the marked suppression of $T_{\rm c}$ and the SC condensation energy $U_0$ in detail within the present scenario in subsection\ \ref{sec:sec34}.
\label{sec:sec32}

\subsection{Superconducting condensation energy of La$_{2-{\it x}}$Sr$_{\it x}$Cu$_{1-{\it y}}$Ni$_{\it y}$O$_4$}
In this subsection, we discuss the SC condensation energy $U_0$ of LSCNO, which was evaluated from the present data of $\gamma_{\rm s}(T, y)$. The condensation energy $U_0$ at $T$=0 K is given by integrating the entropy difference $S_{\rm n}$$-$$S_{\rm s}$ from $T$=0 K to the temperature $T_{\rm sf}$, around which the SC anomaly starts to evolve, slightly higher than $T_{\rm c}$,  
\begin{equation}
U_{0} = \int_0^{T_{\rm sf}}(S_{\rm n}-S_{\rm s})dT, \label{eq:condensationenergy}
\end{equation}
the subscripts ``s'' and ``n'' stand for the SC and hypothetical normal states at $T$$<$$T_{\rm sf}$, respectively. Given both $\gamma_{\rm s}(T)$ and $\gamma_{\rm n}(T)$ as functions of $T$, we can obtain the entropy $S_{\rm s}(T)$ and $S_{\rm n}(T)$ by executing the integration
\begin{equation}
S_{\rm s, n}(T) = \int_0^{T}\gamma_{\rm s, n}(T)dT, \label{eq:gamma}
\end{equation}
and evaluate the condensation energy $U_0$ using Eq.\ \ref{eq:condensationenergy}.\cite{29Loram} In this study, $\gamma_{\rm n}(T)$ was obtained by linear extrapolation of high-temperature data in the normal state down to below $T_{\rm c}$ so as to satisfy the so-called ``entropy balance''; namely, the constraint that both $S_{\rm s}(T)$ and $S_{\rm n}(T)$ values must be equal to each other at $T_{\rm sf}$.\cite{20Momono, 21Matsuzaki}

\begin{figure}
\begin{center}
\includegraphics[scale=.50]{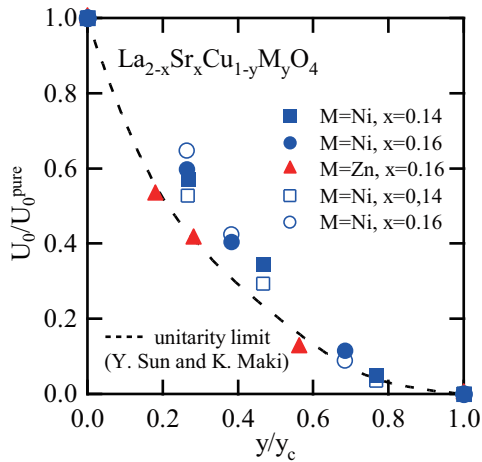}
\end{center}
\caption{(Color online) The SC condensation energy $U_0$ for $\rm La_{2-{\it x}}Sr_{\it x}Cu_{1-{\it y}}Ni_{\it y}O_4$ for $x$=0.14 and $x$=0.16 and $\rm La_{2-{\it x}}Sr_{\it x}Cu_{1-{\it y}}Zn_{\it y}O_4$ for $x$=0.16. Closed symbols are the values obtained from the $C_{\rm el}(T)/T$ vs. $T$ curve, and open symbols are the values calculated by modifying the BCS formula for $U_0$ (Eq.\ \ref{eq:modelcondens}). The dashed line represents the result numerically calculated in the unitarity limit by Sun and Maki.\cite{30Sun}}
\label{fig7}
\end{figure}

The SC condensation energy $U_0$ thus evaluated is plotted as a function of the Ni-concentration $y$ in Fig.\ \ref{fig7}, where $U_0$ and $y$ are normalized with $U_0$ of the pure sample, $U_0^{\rm pure}$, and $y_{\rm c}$ at which the superconductivity is completely suppressed. In Fig.\ \ref{fig7}, $U_0$ of samples doped with nonmagnetic Zn impurities, evaluated as in the case of Ni impurity, is also shown for comparison. The broken line in this figure represents the theoretical result calculated by Sun and Maki for $d$-wave superconductors with impurities (pair-breaking centers) in the unitarity limit, which suppress the SC gap (the SC order parameter) completely at impurity sites.\cite{30Sun} In Zn-doped samples whose SC gap is completely suppressed around the impurities, the experimental values of $U_0$ just fall upon the broken line predicted by the Sun and Maki theory in the unitarity limit (Fig.\ \ref{fig7}), as expected. On the other hand, in Ni-doped samples, the experimental result deviates upward from the theoretical curve, as seen in Fig.\ \ref{fig7}. This fact suggests that Ni impurities will not be in the unitarity limit, and that the suppression of superconductivity will occur predominantly not only around the impurities but uniformly over the whole crystal.\cite{6Ishida, 7Nakano}

\label{sec:sec33}

\subsection{Effective SC gap in La$_{2-{\it x}}$Sr$_{\it x}$Cu$_{1-{\it y}}$Ni$_{\it y}$O$_4$}
In $d$-wave superconductors, it has been broadly thought that impurities will cause pair breaking predominantly in the vicinity of the nodes where the excitation energies of quasiparticles are very small, and that the SC gap would collapse around the nodes. Indeed, in the case of strong impurities in the unitarity limit, whose potential is isotropic and given by the $\delta$-function, Sun and Maki demonstrated that the SC gap will collapse on the FS around the node.\cite{30Sun} Furthermore, when the potential of strong impurities is anisotropic or momentum dependent, it was predicted by Haas {\it et~al.} that the SC gap would also be strongly suppressed around the nodes rather than around the antinodes on the FS.\cite{11Haas} However, such a result seems to be incompatible with the experimental finding that definite nodes remain in the SC gap in LSCNO. On the other hand, Toyama and Ohkawa have discussed the effect of a small number of weak impurities on $d$-wave superconductors on the basis of a self-consistent Born approximation, and shown that the SC gap will be little affected around the nodes as in the present case, whereas it will rather be suppressed around the antinodes through strong scattering of quasiparticles.\cite{31Toyama} However, it seems difficult for this model to reproduce a large recovery of the residual $\gamma$, $\gamma_0(y)$, caused by a small amount of Ni impurity as observed in LSCNO, although the model explains a large reduction of $T_{\rm c}$ well.

In high-$T_{\rm c}$ cuprates, it has been demonstrated that the FS is divided into two segments with different characters. One of them comprises coherent nodal segments of the FS near ($\pm\pi$/2, $\pm\pi$/2), the so-called ``Fermi arc'', on which the $d$-wave SC gap develops below $T_{\rm c}$, and the other incoherent antinodal segments of the FS near ($\pm\pi$, 0) and (0, $\pm\pi$), on which a pseudogap (PG) starts to open above $T_{\rm c}$.\cite{32Norman, 33Tacon, 34Lee, 35Kondo, 36Liu} Recently, it was reported that the gap magnitude at the edges of the coherent nodal FS, $\Delta_{\rm sc}$, dominates $T_{\rm c}$ through BCS (like) relation 2$\Delta_{\rm sc}$=4.3$k_{\rm B}$$T_{\rm c}$; that is, $\Delta_{\rm sc}$ will play a role as the gap maximum $\Delta_0$.\cite{37Yoshida, 38Kurosawa, 39Ido, 40Oda} 

\begin{figure}
\begin{center}
\includegraphics[scale=.50]{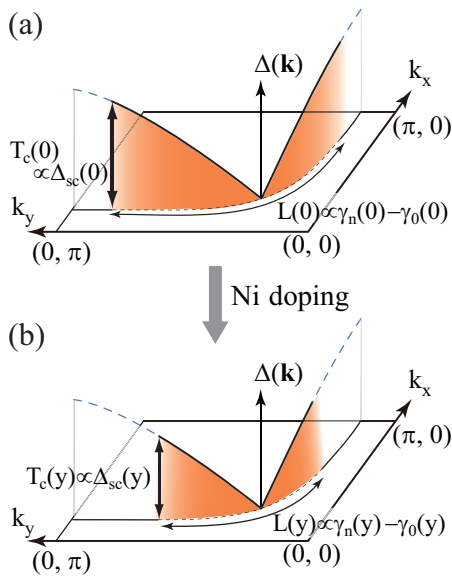}
\end{center}
\caption{(Color online) Schematic illustrations of the SC gap in momentum space for pure and Ni-doped LSCO samples.}
\label{fig8}
\end{figure}

In light of the effective SC gap $\Delta_{\rm sc}$ on the coherent nodal FS, we can speculate the Ni-impurity effect on the SC gap as follows. Ni impurity will predominantly suppress the SC gap on the edges of the coherent nodal FS but in the vicinity of the nodes, as schematically shown in Fig.\ \ref{fig8}. Such a suppression of the SC gap, meaning the shrinkage of the coherent nodal SC segment on the FS, is consistent with the recent report of the ARPES study on Eu-Bi2201\cite{13Okada}. The shrinkage of the nodal SC segment on the FS will bring about a decrease of $\Delta_{\rm sc}$, and consequently suppress $T_{\rm c}$. Since the linear slope of the SC dispersion curve around the nodes decreases only slightly in LSCNO (subsections\ \ref{sec:sec31} and\ \ref{sec:sec32}), the magnitude of $\Delta_{\rm sc}$ linearly depends on the length $L(y)$ of the SC nodal segments on the FS; i.e. $\Delta_{\rm sc}$$\propto$$L(y)$. In LSCNO, $L(y)$ will scale with $\gamma_{\rm n}(y)$$-$$\gamma_0(y)$, $L(y)$$\propto$$\gamma_{\rm n}(y)$$-$$\gamma_0(y)$, as discussed in subsections\ \ref{sec:sec32}. Therefore, taking into account both relations $\Delta_{\rm sc}$$\propto$$L(y)$ and $L(y)$$\propto$$\gamma_{\rm n}(y)$$-$$\gamma_0(y)$, $T_{\rm c}$ is expected to scale with $\gamma_{\rm n}(y)$$-$$\gamma_0(y)$. Actually, this occurred in the present case, as pointed out in subsection\ \ref{sec:sec32} (Fig.\ \ref{fig6}). 

The $y$ dependence of the SC condensation energy at $T$=0 K, $U_0$, can also be explained within the same framework as in the above discussion. According to the BCS theory, the SC condensation energy $U_0$ can be written as follows:
\begin{equation}
U_0 \approx 2.1\times10^{-5}\kappa\gamma_{\rm n}\Delta_0^2/2 \label{eq:modelcondens}
\end{equation}
where prefactor $\kappa$ is $\sim$0.4 for $d$-wave superconductors, and $\Delta_0$ the $d$-wave gap amplitude, $\gamma_{\rm n}$ the normal state $\gamma$ value. In LSCNO, $\gamma_{\rm n}$$-$$\gamma_0$ and $\Delta_{\rm sc}$ can take the place of $\gamma_{\rm n}$ and $\Delta_0$ in Eq.\ \ref{eq:modelcondens} respectively, as mentioned above. In the present study, we estimated the value of $\Delta_{\rm sc}(y)$ from $T_{\rm c}(y)$ using the BCS relation 2$\Delta_{\rm sc}$=4.3$k_{\rm B}T_{\rm c}$, and calculated $U_0$ using estimated values of $\Delta_{\rm sc}(y)$ and experimental values of $\gamma_{\rm n}(y)$$-$$\gamma_0(y)$. The calculated values of $U_0$ are plotted against Ni-concentration $y$ in Fig.\ \ref{fig7} for LSCNO with $x$=0.14 and 0.16. One can see in Fig.\ \ref{fig7} that the calculated $U_0$ values (open symbols) reproduce the experimental ones (closed symbols) of LSCNO well. This finding also supports the present scenario in which the SC gap of LSCNO is locally suppressed around the edge of the coherent nodal FS and the SC gap remains practically intact in the vicinity of the nodes. Such local suppression of the SC gap on the FS will lead to uniform pair breaking in real space, as observed in LSCNO.
\label{sec:sec34}

\section{Summary}
In this study, to elucidate the Ni-impurity effect on the SC gap of high-$T_{\rm c}$ cuprates, we investigated the $H$ dependence and $T$ dependence of the electronic specific heat in $\rm La_{2-{\it x}}Sr_{\it x}Cu_{1-{\it y}}Ni_{\it y}O_4$ (LSCNO). The obtained results can be summarized as follows.

1. The residual $\gamma$ value of LSCNO at $T$=0 K, $\gamma_0$, increases with the application of $H$ along the direction perpendicular to the CuO$_2$ plane. The increase of $\gamma_0$, $\Delta \gamma_0$, follows the Volovik relation $\Delta \gamma_0$=$A\sqrt{H}$, characteristic of $d$-wave superconductivity with line nodes, indicating that definite nodes remain in the SC gap in LSCNO. 

2. In LSCNO, the SC part of $C_{\rm el}/T$, $\gamma_{\rm s}$$\equiv$$C_{\rm el}/T$$-$$\gamma_0$, exhibits a SC anomaly with an abrupt increase at $T_{\rm c}$ and $T$-linear dependence at $T$$\ll$$T_{\rm c}$, although $\gamma_0(y)$ is largely enhanced. The $T$-linear dependence at $T$$\ll$$T_{\rm c}$, characteristic of the existence of the line nodes in the SC gap, is consistent with the results of the magnetic field experiment. Furthermore, it was found for LSCNO that $T_{\rm c}(y)$ scales with the factor $\gamma_{\rm n}(y)$$-$$\gamma_0(y)$. Such results can be explained by a simple model in which Ni impurities suppress the SC gap predominantly at the edges of the coherent nodal FS, leaving the SC gap virtually unchanged in the vicinity of the nodes. 

3. The reduction of $U_0$ in LSCNO is smaller than the theoretical result for impurities in the unitarity limit, which can explain the reduction of $U_0$ for Zn impurity very well. In LSCNO, the result of $U_0$ can be explained as well as $T_{\rm c}$ within the framework of the present scenario. 

\vspace{1\baselineskip}
\begin{center}
{\bf Acknowledgments}
\end{center}
\vspace{1\baselineskip}

We are grateful to S. Oinuma, I. Kawasaki and H. Amitsuka for useful experimental support in the specific heat measurements under magnetic fields. We also thank F. J. Ohkawa for valuable discussions. This work was supported in part by a Grant-in-Aid for Scientific Research and the 21st century COE program ``Topological Science and Technology'' from the Ministry of Education, Culture, Sports, and Technology of Japan.

\end{document}